\documentclass[a4paper,12pt]{article}
\usepackage[dvips]{graphicx}
\usepackage{amsmath}
\begin{document}
\begin{center}
{\large \bf Is there a need for dark matter in galaxies?}
\vspace{5ex}

I. K. Rozgacheva$^a$, A. A. Agapov$^b$
\vspace{2ex}

{\small \it All-Russian Institute for Scientific and Technical Information\\
of Russian Academy of Sciences (VINITI RAS), Moscow, Russia}
\vspace{1ex}

E-mail: $^a$rozgacheva@yandex.ru, $^b$agapov.87@mail.ru
\vspace{4ex}
\end{center}

\begin{abstract}
We present arguments indicating that galaxies and their clusters should be considered as open developing systems. Galaxies interact with the intergalactic medium and are not in a virial equilibrium (which is determined by the gravity and rotation). In this case, the conventional interpretation of the rotation curves of galaxy outer regions outside the visible stellar disk (i.e. the presence of high mass DM haloes) may be erroneous. If there is an accretion of the intergalactic medium in these regions, then the orbital velocities of neutral hydrogen clouds are determined not only by the gravitation of the mass inside their orbits. Galaxy clusters accrete the material (intergalactic gas and galaxies) from filaments of the large-scale structure, at the intersections of which they are located. Only their central regions can approach the virial equilibrium. Therefore, the high velocities of galaxies and the high temperatures of the intergalactic gas in a cluster do not necessarily indicate the presence of a large mass of DM in the cluster.
\end{abstract}
\vspace{1ex}

\vspace{3ex}

\begin{center}
{\bf 1. Introduction}
\end{center}

In modern cosmology, the hypothesis of dark matter (DM) is used in numerical simulations of structure and evolution of galaxies. DM is modeled in the N-body approximation, the hydrodynamic approximation is used for the baryonic component. The quality of the simulations is so high that it can already be compared with the actual experiment \cite{1}. And an interesting trend is discernible: the better simulation of processes in the baryon component (star formation, feedback from stars and activity of galaxy's nucleus upon gas and dust, dynamics and kinematics of stars and gas in the galaxy, interaction with the intergalactic medium), the less important role of the dark matter for physical interpretation of observed statistical patterns of galaxies.

In this paper, we present arguments indicating that galaxies and their clusters should be considered as open developing systems. Galaxies interact with the intergalactic medium and are not in a virial equilibrium (the equilibrium is determined by the gravity and rotation). In this case, the conventional interpretation of the rotation curves of galaxy outer regions outside the visible stellar disk (i.e. the presence of high mass DM haloes) may be erroneous. If there is an accretion of the intergalactic medium in these regions, then the orbital velocities of neutral hydrogen clouds are determined not only by the gravitation of the mass inside their orbits. Galaxy clusters accrete the material (intergalactic gas and galaxies) from filaments of the large-scale structure, at the intersections of which they are located \cite{2}, \cite{3}. Only their central regions can approach the virial equilibrium. Therefore, the high velocities of galaxies and the high temperatures of the intergalactic gas in a cluster do not necessarily indicate the presence of a large mass of DM in the cluster.

The halo masses of DM galaxies are estimated by the method of strong gravitational lensing. Usually, the equilibrium distribution of matter in the lensing galaxy and in its halo is assumed. These estimates, naturally, are consistent with estimates of the rotation curves of galaxies within the limits of measurement errors when using the virial equilibrium. It is clear that the method of gravitational lensing will not be effective without this equilibrium.

The absence of a direct proof for the existence of the DM particles required for the standard cosmological $\Lambda$CDM model \cite{4}, \cite{5} and the arguments considered here lead to the conclusion that there is either little or no dark matter. In this case, the standard model needs to be revised.

In general, the hypothesis of DM was used in cosmology when in the model calculations of 1970s it became clear that in order to form the observed galaxies, clusters and superclusters in the Big Bang theory, initial density perturbations with amplitude ${\displaystyle \frac{\delta\rho}{\rho}\sim 10^{-3}}$ are needed. This amplitude should correspond to the CMB temperature anisotropy with the amplitude
$$
\displaystyle \frac{\delta T}{T}\sim\frac13\frac{\delta\rho}{\rho}
$$
(the Silk effect). However, such CMB temperature anisotropy has not been detected. Therefore cosmologists began to use the hypothesis of DM, which does not interact with radiation. The growth of the DM density perturbations can begin much earlier than in the baryonic medium, and at the time of recombination they can have large amplitudes. Due to gravity, these DM density contrasts attract baryons and generate the baryonic matter density contrasts, sufficient to form the large-scale structure of the Universe.

According to the modern data \cite{6}, the CMB anisotropy does not exceed ${\displaystyle \frac{\delta T}{T}\sim 10^{-5}}$.

If we refuse the hypothesis of the dominant role of DM in the formation of the large-scale structure of the Universe, we will have to solve the "problem of large perturbations in the baryonic matter density and the small CMB anisotropy". A long time ago, one of the authors proposed a mechanism for smoothing the CMB anisotropy on density contrasts due to gravitational lensing \cite{7}, \cite{8}. Now this mechanism is called weak gravitational lensing, examples of its application are described, for example, in \cite{9}, \cite{10}. Below we show that weak gravitational lensing can smooth out the primordial anisotropy and perturbations of the CMB spectrum due to the mixing of null geodesics in space-time and photon momentum in the phase space as they propagate in a weakly inhomogeneous universe. This smoothing can accord a relatively large amplitude of primordial density perturbations with a small amplitude of the observed CMB anisotropy. Note that taking into account the weak gravitational lensing of photon trajectories influences the cosmological interpretation of supernova SNIa observation data \cite{11}, \cite{12}.
\vspace{2ex}

\begin{center}
{\bf 2. Are galaxies in a virial equilibrium?}
\end{center}

Qualitatively, hierarchical models of the galaxies formation with DM halo correctly predict the existence of two main morphological types of galaxies: discs (with star formation) and spheroids (without star formation). Disks are formed due to the continuous accretion of the diffuse gas with the conservation of most of its angular momentum. Spheroids are formed during merges of protogalaxies with low gas content and with effective transfer of angular momentum to the halo. In these models, massive galaxies are formed earlier and faster than low-mass galaxies, and taking into account the feedback from stars and active nuclei makes it possible to reproduce the observed rotation curves and the demography of satellite galaxies in the Local Group.

However, in models with DM and with formation of galaxies in an equilibrium state, the following observational facts cannot be reproduced:

\begin{enumerate}
\item The masses and luminosities of galaxies of the same Hubble type increase with increasing redshift $z$ ("downsizing").
\item The more massive the galaxy, the higher the relative abundance of magnesium to iron in its stars \cite{13}, \cite{14}. (This ratio is an indicator of the star formation rate: the higher it is, the faster and shorter the star-forming epoch. In massive elliptical galaxies this epoch occurred at $z>1.5$ and lasted several hundred million years. Star formation in dwarf galaxies is slow and lasts from the epoch $z\sim 2$ \cite{15}, and the duration of its first stage was not less than several billion years according to the data for the Local Group.) ("downsizing").
\item There is an inconsistency between the model cosmological density of the stellar mass and the star formation rate up to $z\sim 8$ \cite{16}.
\item Low metallicity galaxies have a high content of gas and neutral hydrogen \cite{17}, \cite{18}, \cite{19}, \cite{20}, \cite{21}.
\item Nearby stars of the spectral class G have the same solar metallicity, regardless of their age ("G-dwarf paradox"), i.e. there is no age--metallicity correlation. A similar situation takes place for sub-giants \cite{22}.
\item Lenticular field galaxies have extended gas disks, and according to spectral data, their gas is ionized by shock waves, not by stars, its kinematics is irregular, there are gas layers moving in the direction opposite to rotation, or inclined to the star galaxy's star plane \cite{23}.
\item In the six distant field disk galaxies ($0.6<z<2.6$) with continuing star formation, the "tails of the rotation curves" were not found (observations in the hydrogen emission line), the rotation curves corresponding to the model in which only the baryon disk is present \cite{24}.
\end{enumerate}

The listed facts, apparently, are most simply described in the scenario of the low metallicity intergalactic gas accretion. This gas mixes with interstellar gas and reduces its metallicity, which was created due to the stellar evolution. In new born stars metallicity will be comparable with the metallicity of their predecessors. To maintain star formation for a long time, 8--10 billion years, constant accretion of intergalactic gas is needed \cite{25}, \cite{26}, \cite{27}. In this case, the virial equilibrium could be established only in the central parts of galaxies, but this is impeded by the activity of their nuclei.

The global scaling relationships between the mass and the star formation rate, the gas content, the gas metallicity (thanks to them the term "downsizing" was introduced) can qualitatively be reproduced in modern galaxy formation models in connection with the hypothesis of an approximate equilibrium between the accretion of matter on the galaxy and the outflow of matter from it. However, it is not possible to quantitatively reproduce these statistical dependencies and their change with redshift. This indicates a deviation from the equilibrium between accretion and the outflow of matter in real galaxies. The observational evidence of such a deviation was found in \cite{28}.

The correct reproduction of the structural scaling relationships "mass--size", "mass--rotation speed" and their evolution in discs and spheroids, the evolution of the number densities of these two populations remains an unresolved problem. The simulation shows that dissipative processes in the gas component play a major role in explaining the differences in the structural scaling relationships of spheroids and disks \cite{29}, \cite{30}, rather than the properties and dynamics of DM halo.

The above facts lead us to the conclusion that there is either little or no dark matter. In this case, a rethinking of the standard cosmological model is necessary.
\vspace{2ex}

\begin{center}
{\bf 3. The propagation of photons in a weakly inhomogeneous cosmological model}
\end{center}

The propagation of photons in a weakly inhomogeneous universe was first considered in \cite{31}, \cite{32}. In \cite{31}, the gravitational effect of stationary density perturbations on the propagation of light signals is considered. In this case, only the photon trajectories change (gravitational lensing) and the observed angular size of the photon source changes. In \cite{32}, the influence of nonstationary large-scale inhomogeneities of the gravitational field on the photon frequency without changing their trajectories is considered (the Sachs--Wolfe effect). In this case, the observed brightness of the source changes.

We consider the propagation of photons in a weakly inhomogeneous universe, taking into account the change of their trajectories and frequencies due to nonstationary scalar perturbations of the space-time metric.

For simplicity, we choose a cosmological model with a metric
$$
ds^2=g_{ik}dx^idx^k=a^2\left(d\eta^2-dx^2-dy^2-dz^2\right), \eqno (1)
$$
where the Latin indices run through the values 0, 1, 2, 3; the scale factor $a\sim\eta^2\sim t^{2/3}$; $t$\,is the cosmological time; $\eta$ is the conformal time; $x$, $y$, $z$ are the spatial coordinates. The model is filled with a medium that has a negligible pressure and has density perturbations
$$
\delta_j=\left(\frac{\delta\rho}{\rho}\right)_j\ll 1, \eqno (2)
$$
where $j=1, 2, 3, ...$. It is well known \cite{33} that the considered model has a growing mode of density perturbations: $\delta_j\sim t^{2/3}$.

The medium is transparent for radiation (there is no scattering of photons), and only the gravitational field of density perturbations affects the propagation of photons.

The random spatial distribution of the perturbations (2) is described by the function
$$
f\left(r-r_j\right)=\overline{N}\delta\left(r-r_j\right)+\overline{N}^2\xi\left(r-r_j\right), \eqno (3)
$$
where $\overline{N}$ is the average number of density perturbations in a unit of the comoving volume, $\delta\left(r-r_j\right)$ is the delta function, $\xi\left(r-r_j\right)$ is the correlation function, the perturbation $\delta_j$ is in the neighborhood of the point with coordinates $r_j=\left(x_j, y_j, z_j\right)$.

Density perturbations create scalar perturbations of the metric (synchronous calibration \cite{33}):
$$
g_{ik}\to g_{ik}+h_{ik},
$$
$$
h_{00(j)}=0=h_{0\alpha(j)}, \eqno (4)
$$
$$
h_{\alpha\beta(j)}=\frac13g_{\alpha\beta}\left(\mu_j+\lambda_j\right)-e^{(1)}_{\alpha}e_{\beta (1)}\lambda_{(j)}.
$$
Here the Greek indices run through the values 1, 2, 3 and reference vectors $e_{i(b)}$ are used, and $g_{ik}=e_{i(b)}e^{(b)}_k$, $b=0, 1, 2, 3$, ${e_{i(1)}=\left(0, -a, 0, 0\right)}$, ${e^{(1)}_i=\left(0, a, 0, 0\right)}$. For a growing perturbation mode, the functions $\mu_j$, $\lambda_j$ are:
$$
\mu_j\left(\eta, x^{\alpha}\right)=\int C_j(n)\left(1-\frac{(n\eta)^2}{15}\right)\frac{e^{in_{\alpha}\left(x^{\alpha}-x_j^{\alpha}\right)}}{4\pi\left|r-r_j\right|}d^3n, \eqno (5)
$$
$$
\lambda_j\left(\eta, x^{\alpha}\right)=\int C_j(n)\left(1+\frac{(n\eta)^2}{15}\right)\frac{e^{in_{\alpha}\left(x^{\alpha}-\tilde{x}^{\alpha}\right)}}{4\pi\left|r-r_j\right|}d^3n,
$$
where the functions $C_j(n)$ depend on the initial properties of the perturbations, ${n^2=g_{ik}n^in^k}$, $\left|r-r_j\right|$ is the distance between the points.

For linear perturbations, the superposition principle holds, and therefore the metric in the neighborhood of a point $r$ has the form:
$$
\Delta_{ik}=g_{ik}(\eta)+\sum_{j=1}^Nh_{ik(j)}f(r-r_j). \eqno (6)
$$
Here, $\Delta_{00}=a^2$, $\Delta_{0\alpha}=0$; $N$ is the number of density inhomogeneities in the comoving volume with the center at the point $r$  and the radius of the gravitational interaction $ct$. The space (6) describes the perturbation ensemble $\{\delta_j\}$.

Let us consider light geodesics. The vector $\displaystyle p^i=\frac{q^i}{a^2}$ is an isotropic vector of the metric (1): $p_ip^i=0$, the vector $q^i$ is independent of the coordinates $x^i$,
$$
p_i=g_{ik}p^i=a^2\eta_{ik}p^i=q_i.
$$
The perturbations (4) lead, first, to a change in the modulus $p=\left(p_0p^0\right)^{1/2}$   (the Sachs--Wolfe effect \cite{32}), and, secondly, to the gravitational lensing, i.e. the signal propagation change \cite{31} in the space (1). Let us consider the second effect.

The change in the direction of the geodesic line with the tangent vector $p_i$ will be characterized by the deviation vector of the geodesics $\sigma_i$: $p_i\to p_i+\sigma_i$, $\Delta_{ik}p^i\sigma^k=0$.

From the condition for the light geodesic of space (6)
$$
\Delta_{ik}\left(p^i+\sigma^i\right)\left(p^k+\sigma^k\right)=0,
$$
we find that
$$
\Delta_{ik}\sigma^i\sigma^k=0.
$$
The vector $\sigma_i$ satisfies the equation of geodesic deviation (Jacobi equation):
$$
\frac{D^2\sigma_i}{d\nu^2}=R_{iklm}p^kp^l\sigma^m. \eqno (7)
$$
Here $R_{iklm}$ is the Riemann tensor of the space (6); $\nu$ is the affine parameter of the light geodesic of space (1) with the tangent vector $\displaystyle p^i=\frac{dx^i}{d\nu}$; $\displaystyle \frac{D}{d\nu}$ is the covariant derivative.

The angle of rotation $\phi$ for the vector $p^{\alpha}$ is determined from the expression:
$$
\displaystyle \cos{\phi}=\frac{\left(p^{\alpha}+\sigma^{\alpha}\right)p^{\alpha}}{\sqrt{\displaystyle p^{\alpha}p_{\alpha}\left(p^{\beta}+\sigma^{\beta}\right)\left(p_{\beta}+\sigma_{\beta}\right)}}
\approx
1+\frac{h_{\alpha\beta}p^{\alpha}p^{\beta}}{\displaystyle p^0p_0\left(1+\frac{\sigma^0}{p^0}\right)}. \eqno (8)
$$
As can be seen from the second part of equation (8), the angle of the geodesic deviation depends not only on the metric perturbations, but also on the modulus $p=\left(p_0p^0\right)^{1/2}$. The energy of a photon propagating along a light geodesic is proportional to $p$. It follows from equation (8) that the smaller the photon energy, the greater its geodesic deflection by gravitational perturbations. This effect is important to take into account when assessing the smoothing of the CMB anisotropy due to the density perturbations gravity.

It suffices to analyze the solution of equation (7) in the tangent plane $\{\eta,x\}$, $x=x^1$ of a homogeneous and isotropic space-time with the metric (1).

Let $q^i=\left(q^0, q^1, 0 ,0\right)$, $\sigma_i=\left(\sigma_0, \sigma_1, \sigma_2, 0\right)$. From the conditions $q^iq_i=0$, $q^i\sigma_i=0$, $\sigma^i\sigma_i=0$ and $\sigma^i=\Delta^{ik}\sigma_k$ we find:
$$
q^0=-q^1,\ \ \ q_0=q_1,\ \ \ \sigma_0=\sigma_1,\ \ \ \sigma^1=-\sigma^0+h_1^1\sigma^0-h^{12}\sigma_2,\ \ \ \left(\sigma_2\right)^2\approx h_1^1\left(\sigma_0\right)^2.
$$
Taking into account that for the metric (1)
$$
R_{0101}=-\frac{2a^2}{\eta^2},
$$
from equation (7) one can find the following equation:
$$
\frac{\partial^2\sigma_0}{\partial\eta^2}-2\frac{\partial^2\sigma_0}{\partial\eta\partial x}+\frac{\partial^2\sigma_0}{\partial x^2}=2h_1^1\frac{\sigma_0}{\eta^2}, \eqno (9)
$$
$$
h_1^1=-\left(\epsilon_1+\epsilon_2\eta^2\right)N^2,
$$
$$
\epsilon_1=\frac23\sum_{j=1}^N\xi\int C_j\frac{e^{in_{\alpha}\left(x^{\alpha}-x_j^{\alpha}\right)}}{4\pi\left|r-r_j\right|}d^3n, \eqno (10)
$$
$$
\epsilon_2=\frac1{15}\sum_{j=1}^N\xi\int C_jn^2\frac{e^{in_{\alpha}\left(x^{\alpha}-\tilde{x}^{\alpha}\right)}}{4\pi\left|r-r_j\right|}d^3n.
$$

For the existence of bounded coefficients (10), the correlation function $\xi\left(r-r_j\right)$ should be a decreasing function for positive perturbations
$$
\delta_j=-\frac3{5\kappa a^2\rho c^2}\int C_jn^2e^{in_{\alpha}\left(x^{\alpha}-x_j^{\alpha}\right)}d^3n,
$$
where $\kappa$ is the Einstein gravitational constant, and $\rho$ is the medium density. In this case, for an unchanging number of density inhomogeneities in a unit of comoving volume, the solution of equation (9) has the form:
$$
\sigma_0\approx\sigma_0\left(\eta_0\right)\exp\left[\left(\frac{t}{T}\right)^{1/3}+\left(2\epsilon_2\right)^{1/2}Nx\right], \eqno (11)
$$
where the characteristic time of exponential divergence of the geodesics is equal to
$$
T=\frac2{3H_0}\left[2\epsilon_2\left(1+z_0\right)N^2\right]^{-3/2},
$$
where $H_0$ is the Hubble constant, $z_0$ is the redshift at the initial time $\eta_0$.

Thus, when $\delta_j>0$ the distance between the light geodesics grows exponentially in the collective gravitational field of density inhomogeneities. This field leads to the fact that the direction in which radiation can be detected does not coincide with the initial direction of the radiation. The image of a very distant source of finite sizes can be substantially blurred. This effect should be most important for CMB photons, which propagate in the nonstationary large-scale structure of the Universe.


\vspace{3ex}

\begin{thebibliography}{99}
\bibitem{1} Somerville\;R.\;S., Dav\'e\;R. Physical models of galaxy formation in a cosmological framework. // Annual Review of Astronomy and Astrophysics, 2015, V.\,53, P.\,51-113. (arXiv:1412.2712)
\bibitem{2} Hung\;C.-L., Casey\;C.\;M., Chiang\;Y.-K., Capak\;P.\;L. et\;al. Large-scale structure around a ${\quad z=2.1}$ cluster. // Astrophys.\,J., 2016, V.\,826, N.\,2, id.\,130. (arXiv:1605.07176)
\bibitem{3} Kim\;S., Rey\;S.-C., Bureau\;M., Yoon\;H. et\;al. Large-scale filamentary structures around the Virgo cluster revisited. // Astrophys.\,J., 2016, V.\,833, N.\,2, id.\,207. (arXiv:1611.00437)
\bibitem{4} Liu\;J., Chen\;X., Ji\;X. Current status of direct dark matter detection experiments. // Nature Physics, 2017, V.\,13, P.\,212-216. (arXiv:1709.00688)
\bibitem{5} Zasov\;A.\;V., Saburova\;A.\;S., Khoperskov\;A.\;V., Khoperskov\;S.\;A. Dark matter in galaxies. // Uspekhi Fizicheskikh Nauk, 2017, V.\,187, P.\,3-44.
\bibitem{6} Planck Collaboration, Ade\;P.\;A.\;R., Aghanim\;N., Arnaud\;M., Ashdown\;M. et\;al. Planck 2015 results XIII. Cosmological parameters. // Astronomy \& Astrophysics, 2016, V.\,594, id.\,A13.
\bibitem{7} Rozgacheva\;I.\;K. Is the uniformity of the microwave background radiation an evidence of a homogeneity of the Universe? // Gravity Research Foundation Awards for Essays – 1989, Honorable Mention Essay, [URL: https://static1.squarespace.com/static/5852e579be659442a01f27b8/t/
    5873ea7da5790a59c9157fbf/1483991679256/1989abstracts.pdf ]
\bibitem{8} Rozgacheva\;I.\;K. Is the uniformity of the microwave background radiation an evidence of a homogeneity of the Universe? Proceedings of the Fifth Marcel Grossmann  Meetings on General Relativity and Gravitation, Australia, 1988, P.\,1113-1117.
\bibitem{9} Lewis\;A., Challinor\;A. Weak gravitational lensing of the CMB. // Physics Reports, 2006, V.\,429, N.\,1, P.\,1-65. (arXiv:astro-ph/0601594)
\bibitem{10} Raghunathan\;S., Bianchini\;F., Reichardt\;C.\;L. Imprints of gravitational lensing in the Planck CMB data at the location of WISExSCOS galaxies. // arXiv:1710.09770 (2017).
\bibitem{11} Sereno\;M., Piedipalumbo\;E., Sazhin\;M.\;V. Effects of quintessence on observations of Type\;Ia supernovae in the clumpy universe. // MNRAS, 2002, V.\,335, N.\,4, P.\,1061-1068.
\bibitem{12} Clarkson\;C., Ellis\;G.\;F.\;R., Faltenbacher\;A., Maartens\;R., et\;al. (Mis)interpreting supernovae observations in a lumpy universe. // MNRAS, V.\,426, N.\,2, P.\,1121-1136. (arXiv:1109.2484)
\bibitem{13} Mannucci\;F., Cresci\;G., Maiolino\;R., Marconi\;A. et\;al. A fundamental relation between mass, star formation rate and metallicity in local and high-redshift galaxies. // MNRAS, 2010, V.\,408, N.\,4, P.\,2115-2127. (arXiv:1005.0006)
\bibitem{14} Lara-L\'opez\;M.\;A., Cepa\;J., Bongiovanni\;A., P\'erez Garc\'ia\;A.\;M. et\;al. A fundamental plane for field star-forming galaxies. // Astronomy \& Astrophysics, 2010, V.\,521, id.\,L53. (arXiv:1005.0509)
\bibitem{15} Geha\;M., Blanton\;M.\;R., Yan\;R., Tinker\;J.\;L. A stellar mass threshold for quenching of field galaxies. // Astrophys.\,J., 2012, V.\,757, N.\,1, id.\,85. (arXiv:1206.3573)
\bibitem{16} Yu\;H., Wang\;F.\;Y. On the inconsistency between cosmic stellar mass den¬sity and star formation rate up to ${\quad z\sim 8}$. // Astrophys.\,J., 2016, V.\,820, N.\,2, id.\,114. (arXiv:1602.01985)
\bibitem{17} Peeples\;M.\;S., Shankar\;F. Constraints on star formation driven galaxy winds from the mass-metallicity relation at ${\quad z=0}$. // MNRAS, 2011, V.\,417, N.\,4, P.\,2962-2981.
\bibitem{18} Lara-L\'opez\;M.\;A., Hopkins\;A.\;M., L\'opez-S\'anchez\;A.\;R., Brough\;S. et\;al. Galaxy And Mass Assembly (GAMA): the connection between metals, specific SFR and H\,I gas in galaxies: the Z-SSFR relation. // MNRAS, 2013, V.\,433, N.\,1, P.\,L35-L39. (arXiv:1304.3889)
\bibitem{19} Lara-L\'opez\;M.\;A., Hopkins\;A.\;M., L\'opez-S\'anchez\;A.\;R., Brough\;S. et\;al. Galaxy And Mass Assembly (GAMA): a deeper view of the mass, metallicity and SFR relationships. // MNRAS, 2013, V.\,434, N.\,1, P.\,451-470. (arXiv:1306.1583)
\bibitem{20} Bothwell\;M.\;S., Smail\;I., Chapman\;S.\;C., Genzel\;R. et\;al. A survey of molecular gas in luminous sub-millimetre galaxies. // MNRAS, 2013, V.\,429, N.\,4, P.\,3047-3067. (arXiv:1205.1511)
\bibitem{21} Bothwell\;M.\;S., Maiolino\;R., Kennicutt\;R., Cresci\;G. et\;al. A fundamental relation between the metallicity, gas content and stellar mass of local galaxies. // MNRAS, 2013, V.\,433, N.\,2, P.\,1425-1435. (arXiv:1304.4940)
\bibitem{22} Nordstr\"om\;B., Mayor\;M., Andersen\;J., Holmberg\;J. et\;al. The Geneva-Copenhagen survey of the Solar neighbourhood. Ages, metallicities, and kinematic properties of ${\sim14\,000}$ F and G dwarfs. // Astronomy \& Astrophysics, 2004, V.\,418, P.\,989-1019. (arXiv:astro-ph/0405198)
\bibitem{23} Katkov\;I.\;Y., Kniazev\;A.\;Y., Sil'chenko\;O.\;K. Kinematics and stellar populations in isolated lenticular galaxies. // Astronom.\,J., 2015, V.\,150, N.\,1, id.\,24. (arXiv:1505.01386)
\bibitem{24} Genzel\;R., F\"orster\;Schreiber\;N.\;M., \"Ubler\;H., Lang\;P. et\;al. Strongly baryon-dominated disk galaxies at the peak of galaxy formation ten billion years ago. // Nature, 2017, V.\,543, N.\,7645, P.\,397-401. (arXiv:1703.04310)
\bibitem{25} Dav\'e\;R., Oppenheimer\;B.\;D., Finlator\;K. Galaxy evolution in cosmological simulations with outflows – I. Stellar masses and star formation rates. // MNRAS, 2011, V.\,415, N.\,1, P.\,11-31. (arXiv:1103.3528)
\bibitem{26} Dav\'e\;R., Finlator\;K., Oppenheimer\;B.\;D. Galaxy evolution in cosmological simulations with outflows – II. Metallicities and gas fractions. // MNRAS, 2011, V.\,416, N.\,2, P.\,1354-1376. (arXiv:1104.3156)
\bibitem{27} Saintonge\;A., Kauffmann\;G., Wang\;J. Kramer\;C. et\;al. COLD GASS, an IRAM legacy survey of molecular gas in massive galaxies – II. The non-universality of the molecular gas depletion time-scale. // MNRAS, 2011, V.\,415, N.\,1, P.\,61-76. (arXiv:1104.0019)
\bibitem{28} Bouch\'e\;N., Finley\;H., Schroetter\;I., Murphy\;M.\,T. et\;al. Possible signatures of a cold-flow disk from MUSE using a ${\quad z\sim 1}$ galaxy-quasar pair toward SDSS J1422–0001. // Astrophys.\,J, 2016, V.\,820, N.\,2, id.\,121. (arXiv:1601.07567)
\bibitem{29} Covington\;M.\;D., Primack\;J.\;R., Porter\;L.\;A., Croton\;D.\;J. et\;al. The role of dissipation in the scaling relations of cosmological merger remnants. // MNRAS, 2011, V.\,415, N.\,4, P.\,3135-3152. (arXiv:1101.4225)
\bibitem{30} Porter\;L.\;A., Somerville\;R.\;S., Primack\;J.\;R., Johansson\;P.\;H. Understanding the structural scaling relations of early-type galaxies. // MNRAS, 2014, V.\,444, N.\,1, P.\,942-960. (arXiv:1407.0594)
\bibitem{31} Gunn\;J.\;E. A fundamental limitation on the accuracy of angular measurements in observational cosmology. // Astrophys.\,J., 1967, V.\,147, P.\,61-72.
\bibitem{32} Sachs\;R.\;K., Wolfe\;A.\;M. Perturbations of a cosmological model and angular variations of the microwave background. // Astrophys.\,J., 1967, V.\,147, P.\,73-90.
\bibitem{33} Landau\;L.\;D., Lifshitz\;E.\;M. The Classical Theory of Fields. Vol.\,2 (4th\,ed.), 1975. Butterworth-Heinemann. ISBN 978-0-7506-2768-9.
\end{thebibliography}
\end{document}